# The MICA Experiment: Astrophysics in Virtual Worlds

## S. G. Djorgovski

California Institute of Technology, MS 249-17, Pasadena, CA 91125, USA. Email: george@astro.caltech.edu.

## Piet Hut

Institute for Advanced Study, 1 Einstein Drive, Princeton, NJ 08540. Email: piet@ias.edu.

## Rob Knop

Quest Univ. Canada, 3200 University Bvd., Squamish, BC, Canada V8B 0N8. Email: rknop@pobox.com.

## Giuseppe Longo

Dept. of Physical Sciences, Univ. Federico II, Via Cintia 9, 80126 Napoli, Italy. Email: longo@na.infn.it.

## Steve McMillan

Physics Dept., Drexel University, Philadelphia, PA 19104, USA. Email: steve@physics.drexel.edu.

## Enrico Vesperini

Physics Dept., Drexel University, Philadelphia, PA 19104, USA. Email: vesperin@physics.drexel.edu.

## Ciro Donalek

California Institute of Technology, MS 249-17, Pasadena, CA 91125, USA. Email: donalek@astro.caltech.edu.

## Matthew Graham

California Institute of Technology, MS 158-79, Pasadena, CA 91125, USA. Email: mjg@cacr.caltech.edu.

## Asish Mahabal

California Institute of Technology, MS 249-17, Pasadena, CA 91125, USA. Email: aam@astro.caltech.edu.

## Franz Sauer

California Institute of Technology, MS 249-17, Pasadena, CA 91125, USA. Email: fsauer@caltech.edu.

## Charles White

Jet Propulsion Laboratory, Pasadena, CA 91109, USA. Email: charles.p.white@jpl.nasa.gov.

## Crista Lopes

Dept. of Informatics, University of California, Irvine, CA 92697, USA. Email: lopes@ics.uci.edu.

We describe the work of the Meta-Institute for Computational Astrophysics (MICA), the first professional scientific organization based in virtual worlds. MICA was an experiment in the use of this technology for science and scholarship, lasting from the early 2008 to June 2012, mainly using the *Second LIfe*[TM] and *OpenSimulator* as platforms. We describe its goals and activities, and our future plans. We conducted scientific collaboration meetings, professional seminars, a workshop, classroom instruction, public lectures, informal discussions and gatherings, and experiments in immersive, interactive visualization of high-dimensional scientific data. Perhaps the most successful of these was our program of popular science lectures, illustrating yet again the great potential of immersive VR as an educational and outreach platform. While the members of our research groups and some collaborators found the use of immersive VR as a professional telepresence tool to be very effective, we did not convince a broader astrophysics community to adopt it at this time, despite some efforts; we discuss some possible reasons for this non-uptake. On the whole, we conclude that immersive VR has a great potential as a scientific and educational platform, as the technology matures and becomes more broadly available and accepted.

## Introduction

Virtual Worlds (VWs) and immersive Virtual Reality (VR) technologies are still in their infancy, and yet they hold a huge transformative potential. They may presage the emerging "3D Web", that may be as transformative as the WWW itself, if not more. This includes their possible uses in science, scholarship, and education.

We face a dual problem of engaging the broad academic community in their use and exploration for scientific and scholarly research in general, and at the same time taping into its innovation potential to help shape and develop the VWs and enhance their utility and functionality. It would be healthy to have the intellectual leadership and rigor in this arena come form the academia, rather than from the games industry alone. Yet, the scientific community at large seems to be largely unaware of this technological emergence or its potential.

While there has been a slowly growing interest in VWs and engagement of the academic community in the humanities and social sciences (e.g., Bainbridge 2007, 2010), with a few exceptions (e.g., Lang & Bradley 2009) the "hard sciences" community has yet to engage meaningfully in these interesting, possibly transformative developments. Aside from being insufficiently informed, and the natural inertia in adopting radically new things, one reason for this negligence may be a lack of the real-life examples of the scientific utility of VWs.

With this in mind, we formed the Meta-Institute for Computational Astrophysics (MICA; http://mica-vw.org), to the best of our knowledge the first professional scientific organization based exclusively in virtual worlds (VWs). Our goals were to explore the utility of the emerging VR and VWs technologies for scientific and scholarly work in general, and to facilitate and accelerate their adoption by the scientific research community.

The charter goals of MICA were:

- Exploration, development and promotion of VWs and VR technologies for professional research in astrophysics and related fields.

- To provide and develop novel social networking venues and mechanisms for scientific collaboration and communications, including professional meetings, effective telepresence, etc.

- Use of VWs and VR technologies for education and public outreach.

- To act as a forum for exchange of ideas and joint efforts with other scientific disciplines in promoting these goals for science and scholarship in general.

MICA was formed in the early 2008, following the early explorations by Hut (2006, 2008), and lasted until June 2012. Its work continues under the auspices of the Caltech Astroinformatics group. Some of our work to date has been described in McMillan et al. (2009), Knop et al. (2010), and Djorgovski et al. (2010a,b), and in a number of other conference presentations.

While our initial activities were conducted in the VW of *Qwaq* (renamed since to *TelePlace*), we quickly migrated to *Second LIfe*[TM] (SL) and, more recently, to *OpenSimulator* (*OpenSim*) platforms. SL provided a convenient, well established virtual environment, and the ready audiences for our outreach activities.

Astronomy has been at a leading edge of the e-Science and Cyber-Infrastructure developments, e.g., with the Virtual Observatory framework: a Web-based, distributed research environment for astronomy with massive and complex data sets (see, e.g., Brunner et al. 2001, or Djorgovski & Williams 2005); however that "virtual" is not yet related in any way to the immersive VR or VWs, and is today mainly providing a global data grid of astronomy, with some data services. We also have many on-line forums for research collaborations, such as MODEST (http://www.manybody.org/modest), in which a number of the founding members of MICA were engaged. An emerging discipline of Astroinformatics aims to develop deeper and broader connections between astronomy and applied computer science and information technology (see, e.g., http://astroinformatics2010.org).

MICA was an experiment in academic and scientific practices enabled by the immersive VR technologies, an example of the e-Science, or the "Fourth Paradigm" (Hey et al. 2009; the first three being experiment, analytical theory, and numerical simulations) – a segment in the evolving landscape of computationally enabled science in the 21st century.

## VWs as a Scientific/Scholarly Platform

As most people who have seriously tried them know, VWs are clearly a powerful scientific communication and collaboration platform. In addition to the traditional uses, such as the discussion, conference, or collaboration group discussion venues, there is another important aspect where VWs can play an essential facilitating role:

Genuine interdisciplinary cross-fertilization is a much-neglected path to scientific progress. Given that many of the most important challenges facing us (e.g., the global climate change, energy, sustainability, etc.) are fundamentally interdisciplinary in nature, and not reducible to any given scientific discipline (physics, biology, etc.), the lack of effective and pervasive mechanisms for establishment of inter-, multi-, or cross-disciplinary interactions is a serious problem which affects us all. Engaging in effective interdisciplinary activities requires *easy* and *effective* communication venues, intellectual melting pots where such encounters can occur and flourish. VWs as scientific interaction environments offer a great new opportunity to foster interdisciplinary meetings of the minds. They are easy, free, do not require travel, and the social barriers are very low and easily overcome (the ease and the speed of striking conversations and friendships is one of the more striking features of VWs).

Another interesting question is how immersive VR can be used as a part of novel forms of scientific publishing, either as an equivalent of the current practice of supplementing traditional papers with on-line material on the Web, or even as a *primary* publishing medium. Just as the Web offers new possibilities and modalities for scholarly publishing which do not simply mimic the age-old printed-paper media publishing, so we may find qualitatively novel uses of VWs as a publishing venue in their own right.

Immersive VR environments open some intriguing novel possibilities in the ways in which scientists can set up, perform, modify, and examine the output of numerical simulations. In MICA, we used as our primary science environment the gravitational N-body problem, since that is where our professional expertise is concentrated, but we expect that most of the features we developed will find much broader applicability in the visualization of more general scientific or abstract data sets in arbitrary VW environments.

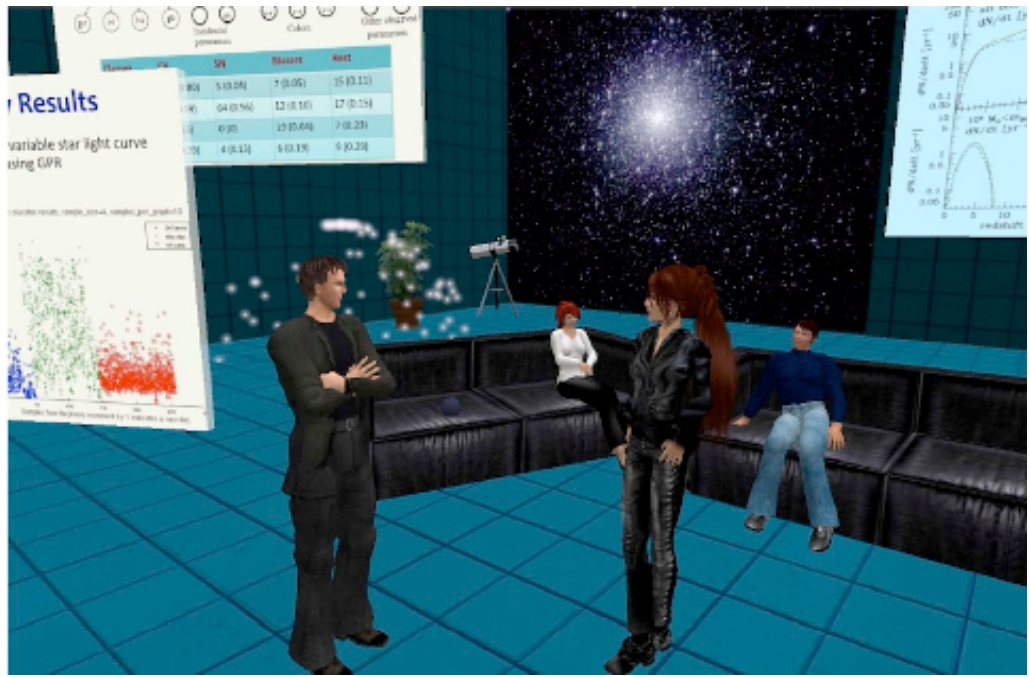

Figure 1: MICA members conducting scientific and collaborative discussions in an immersive environment in SL. Graphics (diagrams, slides, etc.) are imported as textures and displayed on suitable prims.

In a more general context, VWs offer intriguing new possibilities for scientific visualization or "visual analytics". As the size, and especially the *complexity* of scientific data sets increase, effective visualization becomes a key need for data analysis: it is a bridge between the quantitative information contained in complex scientific measurements, and the human intuition which is necessary for a true understanding of the phenomena that are being studied.

Moreover, the human visual perception system is naturally optimized for 3D: we are meant to interact with each other, with objects, and with informational constructs in 3D; the traditional 2D paper or screen paradigm is simply a historical and technological artifact. Perhaps this is the main reason for the "unreasonable effectiveness" of VWs (given the technology's nascent sta) in creating a subjective feeling of a real presence.

VWs provide an easy, portable and inexpensive (or free) venue for a multi-dimensional data visualization, but with an added benefit of being able to interact with the data and with your colleagues, in a truly explorative and collaborative manner.

One increasingly plausible vision of the future is that there will be a synthesis of the Web, with its all-encompassing informational content, and the immersive VR as an interface to it, since it is so well suited to the to the human sensory input mechanisms. We can think of immersive VR as the next generation browser technology, which will be as qualitatively different from the current, flat desktop and web page paradigm, as it was different from the older, terminal screen and file directory paradigm for information display and access. A question then naturally arises: what will be the newly enabled ways of interacting with the informational content of the 3D Web, and how should we structure and architect the information so that it is optimally displayed and searched under the new paradigm?

## Immersive VR as a Scientific Collaboration and Telepresence Platform

We have been using SL extensively as a venue for research group and collaboration discussions, including collaborators worldwide (Fig. 1). Many of us prefer this mode of interaction to Skype or telecon meetings, including the standard videoconferencing. Several research papers have been conceived at these meetings.

Much of our early effort was focused on the visualization and exploration of numerical stellar dynamics simulations in VWs. Throughout 2009, we organized weekly meetings in the MICA building, on our *StellaNova* island, in order to discuss the use of N-body simulations in SL and in *OpenSim*. The meetings were highly successful, and we had a steady audience of ten to twenty participants, partly professional astrophysicists, partly amateur astronomers and others interested in learning more about the gravitational N-body problem and its applications in stellar dynamics simulations of star clusters and galaxies.

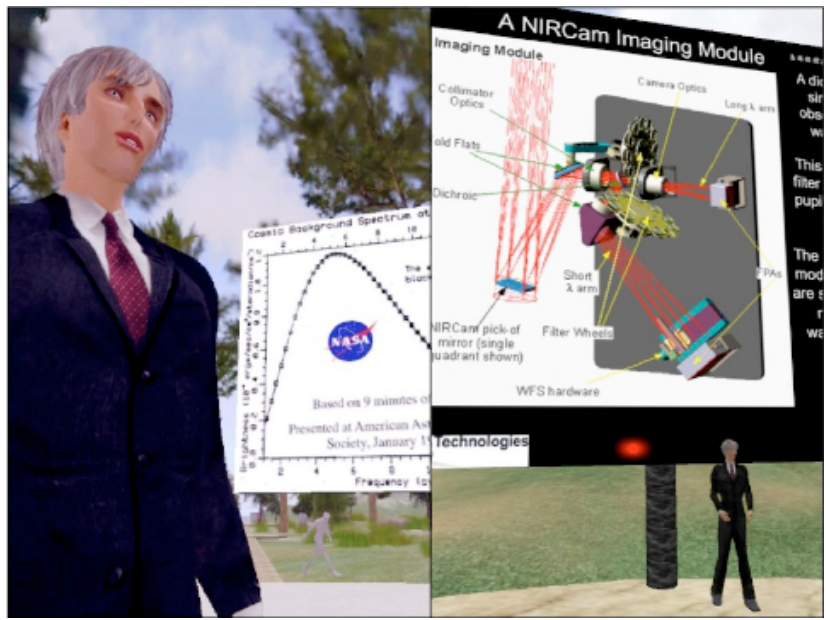

Figure 2: Nobel laureate John Mather (now the Project Scientist for the James Webb Space Telescope) giving one of the MICA technical seminars.

Throughout 2009, we organized weekly meetings in the MICA building on our *StellaNova* island in SL, in order to discuss the use of N-body simulations in SL and in *OpenSim*. The meetings were highly successful, and we had a steady audience of ten to twenty participants, partly professional astrophysicists, partly amateur astronomers and others interested in learning more about the gravitational N-body problem and its applications in stellar dynamics simulations of star clusters and galaxies.

In the first year, we used SL as a platform for our weekly professional seminar series, with a typical attendance of ~ 30 avatars. These seminars served as a device to introduce our colleagues to VWs: we invite them to give a talk as an introductory experience with this medium. One of our speakers (both for a professional seminar, and for a public lecture) was John Mather, a Nobel laureate in physics, and currently the Project Scientist for the James Webb Space Telescope (Fig. 2). This indicates the level of seriousness and the perceptions of at least some members of the professional astrophysics community.

The great majority of those who accept our invitation to speak at the MICA seminars in SL found the experience to be interesting and rewarding. However, while some of the colleagues we attracted in this way remained active in exploring the scientific uses of VWs, the majority did not. We have thus stopped this seminar series, and had seminars on an ad hoc basis, when an appropriate speaker was available.

The lesson learned from this experiment is that the great majority of our colleagues are still leery and reluctant to embrace VWs as a scholarly platform. We consider below some possible reasons for this slow uptake of a highly promising technology. Unfortunately, this is not an unusual situation with the process of the academic community adoptions of any new technology, especially in the internet era, and given the endemic inertia of the academic institutions in adopting new ways of doing business. Some persistence is needed, as well as tangible demonstrations of the utility of these technologies for the scholarly work.

We also conducted a 1-day international workshop on the scientific uses of VWs within SL. This confirmed our expectations that immersive environments represent an effective, easy, inexpensive, and environment-friendly (due to the absence of a physical travel) venue for professional meetings. This was of course already realized by many other groups in the business world and by some government agencies, but it has not yet registered effectively in the academic community at large.

By eliminating the necessity of a physical travel, virtual meetings represent a very "green" technology. While in the early days of the internet there were high expectations for telecommuting, they were dashed by the lack of immediate and subjectively personal interactions (the "watercooler effect"). VWs solve this problem, and we expect that they will have a major impact in this arena, once this technology becomes more broadly accepted.

### Education and Public Outreach

There is of course an extensive literature on the education in VWs, the review of which is beyond the scope of this paper. Gauthier (2007) describes some early astronomy outreach efforts in SL.

We have experimented with a normal classroom instruction in a VW environment (Fig. 3). We plan a more extensive use of VWs for both classroom-style lectures and informal student-faculty discussions.

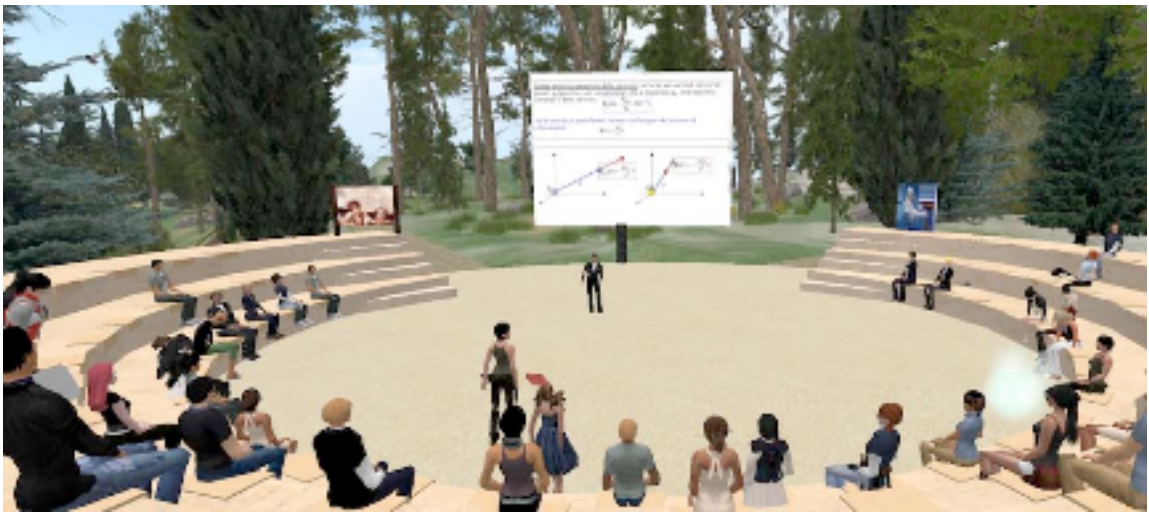

Figure 3: A lecture in an introductory physics class conducted in SL, while the instructor (Prof. G. Longo) was on one continent, and the students on another. The students' reaction to this novel approach was largely very favorable.

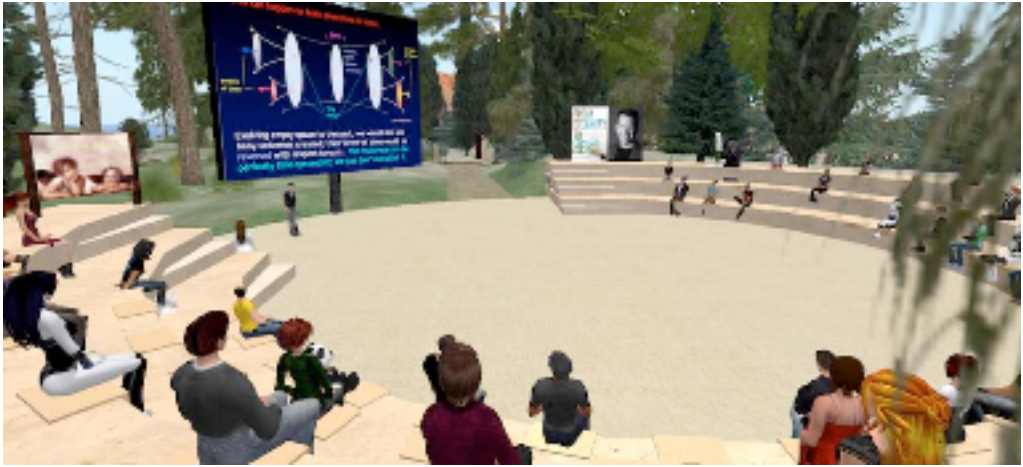

Figure 4. One of the MICA popular lectures: Dr. Sean Carroll from Caltech is explaining the material from his book, *From Eternity to Here*.

We have established a strong and successful program of public lectures (Fig. 4), initially on a bi-weekly, but then on a weekly basis, during the academic terms. These included external speakers, as well as the members of our team. They were very popular, with a typical attendance of ~ 50 – 70 avatars. The slides shown, and in most cases also the audio and/or video (machinima) recordings are posted on the MICA website, and are freely available.

During the first few months, we have also held weekly informal "Ask an Astronomer" sessions, where anyone could ask astronomy or general science questions from one of our professional members. These sessions were very popular with science enthusiasts.

We have started to develop a virtual educational laboratory content, starting with a simple physics experiment, a 3-body gravitational interaction (Fig. 5). Students can modify the initial condition and observe the changes in the outcome, computed in real time. We envision such virtual teaching laboratories as a great potential resource for the schools who cannot afford actual real-life laboratories. Moreover, some experiments simply cannot be done in a real lab context – dynamical evolution of stellar systems being an example. We can develop a virtual lab exercise where one can change the actual physics (e.g., "what if the gravitational force was inversely proportional to the cube of the distance?"), and see the difference in the outcomes.

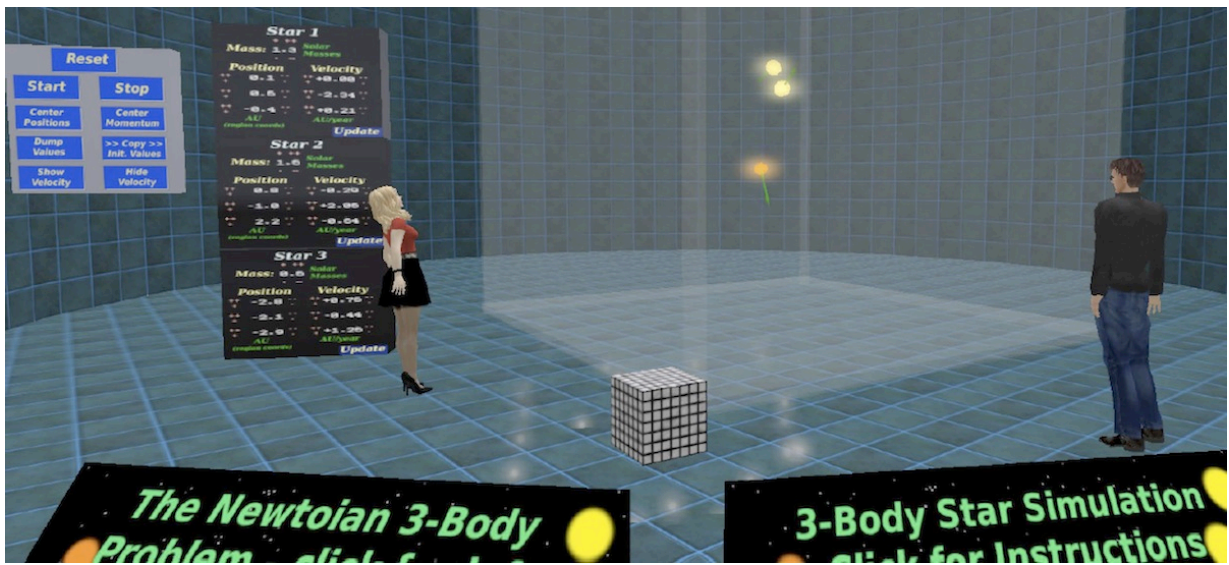

Figure 5. A working demo of a gravitational 3-body problem, deployed at the MICA sim in SL, and developed by Prof. Rob Knop, one of our core team members. This is an example of an interactive, hands-on experiment that a virtual teaching laboratory might contain.

One important feature of VWs is that they lower the social barriers in most human interactions, and education is no exception. People who attended our popular lectures and informal discussions would generally not attempt (or not even have an opportunity) to make comparable contacts in real life. This leveling of an educational playing field may have a huge, beneficial social impact.

## Scientific Data Visualization

Immersive visualization of complex data spaces is now the main research direction we are pursuing. VWs offer intriguing new possibilities for scientific visualization or "visual analytics". As the size, and especially the *complexity* of scientific data sets increase, effective visualization becomes a key need for data analysis: it is a bridge between the quantitative information contained in complex scientific measurements, and the human intuition which is necessary for a true understanding of the phenomena in question. The advantages of VWs in this arena are that the *visual exploration can be collaborative*, as researchers interact with each other at the same time as they interact with the data. It is also a low-cost, highly portable alternative to many other methods of 3D data visualization (e.g., caves, use of special theaters, helmets, or goggles, etc.).

Our initial experiments with immersive visualization of stellar dynamics simulation (Fig. 6) have been described by Farr et al. (2009) and Nakasone et al. (2009). In addition to visualization of pre-made, stationary data sets, we think of visualizing output of numerical simulations or data streams in real time, allowing scientists to interact with their experiment itself as it is ongoing.

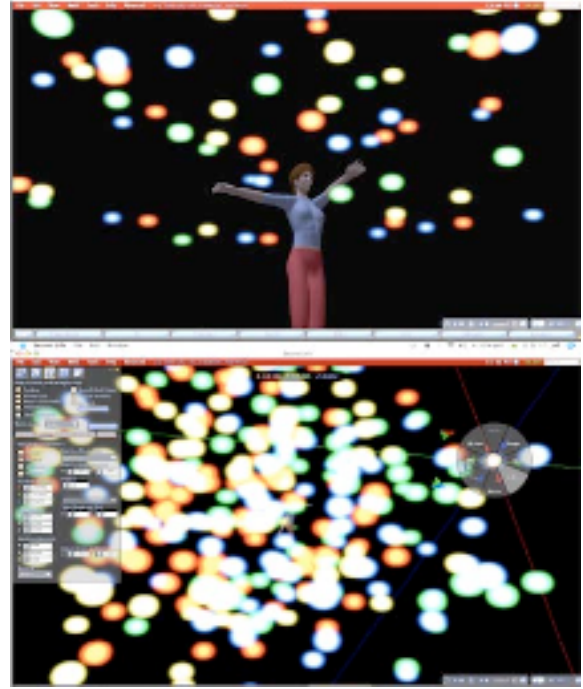

Figure 6. An early example of an immersive visualization of an output of a dynamical simulation of a star cluster, with the scientist interacting with the simulation from within the VW.

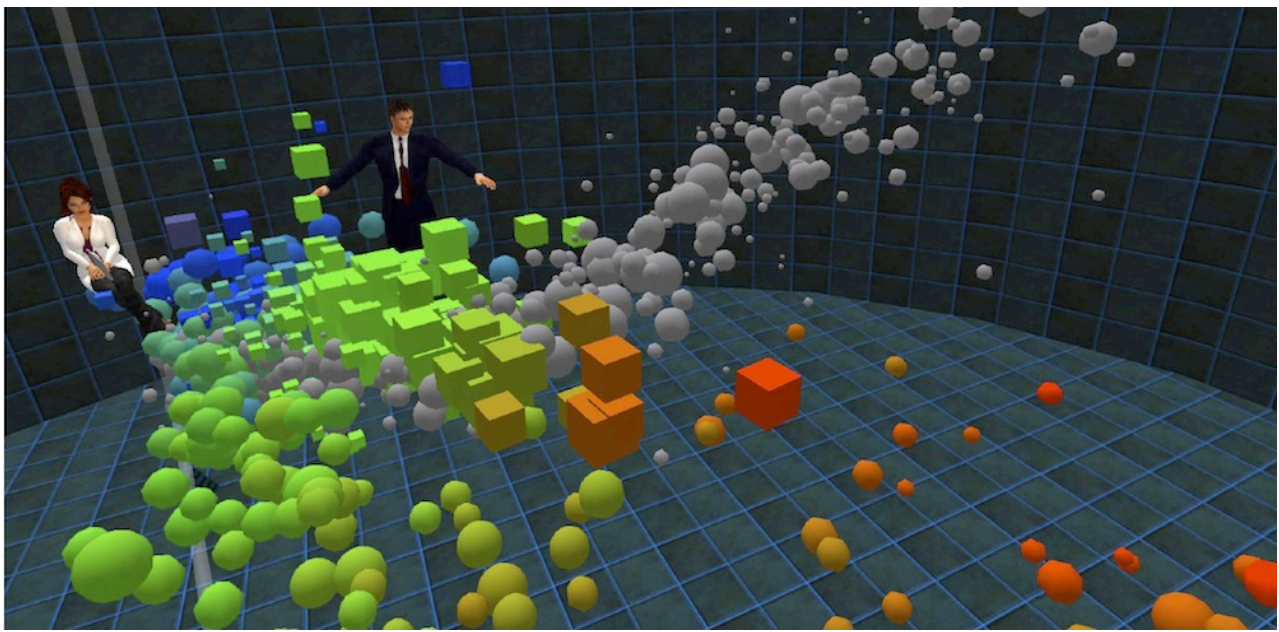

Figure 7. An example of an immersive, collaborative data visualization in SL. The data represent properties of stars, galaxies and quasars from the SDSS sky survey. The XYZ positions, point shapes, sizes, and colors encode different observed parameters.

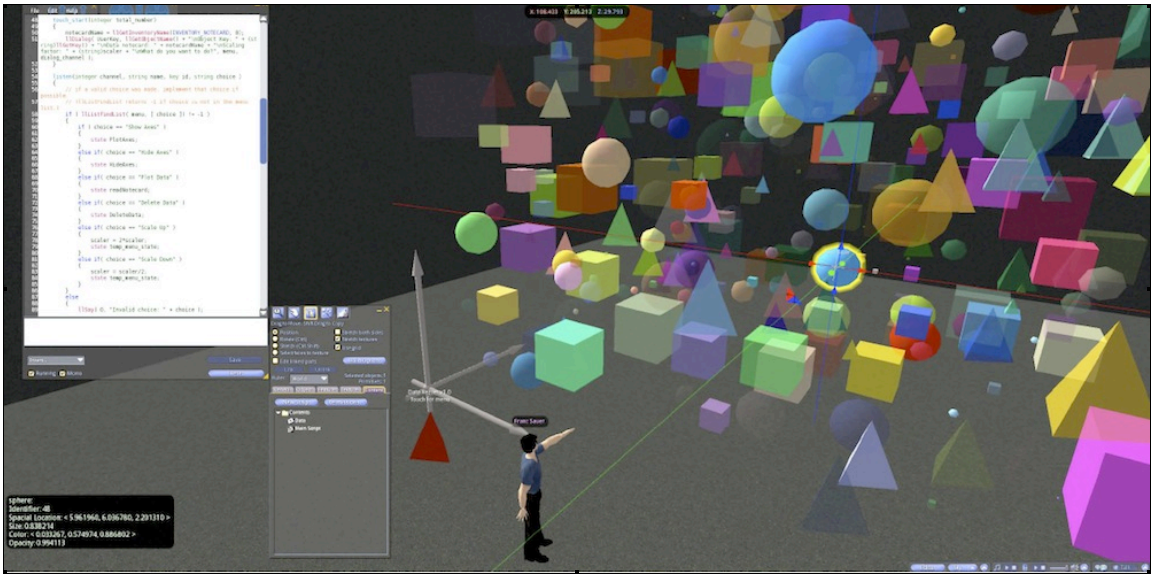

Figure 8. A Caltech student (F. Sauer) experimenting with data visualization scripts in an *OpenSim* world.

In collaboration with Desdemona Enfield (her SL *nom de pixel*), we have developed universal scripts for immersive visualization of highly dimensional data sets. We are using data object shapes, textures, orientations, transparency, rotation, pulsation, etc., to encode additional data dimensions beyond the obvious spatial XYZ coordinates, color, and size of data objects (Fig. 7). This was expanded by us to include the transparency (the alpha layer) as the means of encoding an additional dimension. We can add additional dimensions through the use of object shapes, orientations (for the non-spherical data points), textures, and glyphs. In all, we expect that about a dozen parameter space dimensions can be encoded in these immersive, peudo-3D displays.

An additional functionality we added is the ability to link data objects with the external catalog or database information, e.g., using a simple point-and-click. This information can be displayed using a Web browser window, either external or internal to the VW browser.

We have ported these scripts into the *OpenSim* worlds, initially Intel's *ScienceSim,* and now the *Virtual Caltech* (vCaltech), where the experiments continue. This enabled us to overcome a major obstacl, the limited prim quotas in SL. In our *OpenSim* experiments, we can easily visualize ~ 100,000 data points. Beyond that, individual data point representations are better replaced by isodensity surfaces, at least for the majority of the data; outliers are still best represented as individual data points.

We are now also developing data visualization using the Unity 3D platform, and experimenting with Microsoft's Kinect device as a haptic interface.

## Concluding Comments

MICA was a new type of scientific institution, dedicated to an exploration of immersive VR and VWs technologies for science, scholarship, and education, aimed primarily at physical and other natural sciences. It was an experiment in the new ways of conducting scholarly work, as well as a testbed for new ideas and research modalities.

The central idea here is that immersive VR and VWs are potentially transformative technologies on a par with the Web itself, which can and should be used for serious purposes, including science and scholarship; they are not just a form of games, and that message has to be absorbed by the academic community at large.

Our goal was to engage a much broader segment of the academic community in utilizing, and developing further these technologies. This, in turn, would bring in the new creative potential of the community in developing further the VR and VW technologies themselves.

MICA was intended to be a gateway for other scholars, new to VWs, to start to explore their potential and the practical uses in an easy, welcoming, and collegial environment. However, we did not succeed in engaging a broader segment of the astrophysics community in the adoption (let alone development) of these technologies; the same applies in most other sciences or academic domains. Why are academics so slow to recognize the utility and the potential of VWs and immersive VR?

We have polled our professional members (about 50) who have not continued to use VWs beyond their initial visit as to why. The majority answer was that they did fid the technology interesting, but simply do not have the time to invest in exploring it now.

A part of the answer is that many people (in particular those older than the growing generations of "digital natives") are not used to (and some are simply repelled by) the avatar representations of themselves and their colleagues – it looks like a video game, and not a serious professional activity. An additional factor is the (well deserved) iffy reputation of VWs such as SL and what goes on in them. This stigma must be overcome, if we are to attract a broader community of academic professionals to these technologies, both as users and as developers.

One possible approach is to introduce our skeptical colleagues to VWs through dedicated academic VWs with controlled access. *OpenSim* worlds are currently perhaps the best option in this regard. The SL/*OpenSim* will most likely not be the architecture of the future "3D Web". However, it is a good interim platform to get used to the immersive VR experience and start developing the novel tools for science, scholarship, and education.

Another factor in the slow uptake may be the quality of the user experience in VWs such as SL. Pure commercial games, for example, have vastly better graphics, although they are more rigid, and mostly not user-programmable. Yet there is no obvious alternative to SL and *OpenSim* at this time as scholarly VR experimentation platforms.

We are currently witnessing a dramatic growth in 3D display technologies, largely driven by the entertainment industry. These commercial interests are funding the technology developments in a way that the academic community could not, and that gives us a powerful leverage – just as it happened in the past with computing and information technology.

The lack of an effective uptake of VWs in the academic community is symptomatic of a broader problem: while the information technology evolves on a Moore's law time scale (i.e., a couple of years at most), humans learn new skills and change their behavior on much longer time scales; they simply cannot keep up with the pace of the technology. An even more insidious problem is that academia as an institution evolves even slower, on time scales of decades or centuries. While the new generations of digital natives may find these technologies to be perfectly natural and a standard part of their lives, both personal and professional, we have to accelerate their adoption in the academic and research contexts.

This evolutionary process may have an impact well beyond academia, as these technologies will change the ways we interact, both with each other and with the informational content in the cyberspace. Engaging the academic community in the extensive use of VR and VWs may also lead to novel practical and commercial applications and development directions which we cannot even anticipate today. If immersive VR becomes a major feature of modern society, in commerce, entertainment, etc., the potential impact will be very significant.

## ACKNOWLEDGMENTS

MICA was supported by the NSF grant HCC-0917814. We are grateful to a number of colleagues and volunteers, and in particular: Simon Portegies Zwart, Will Farr, Cassandra Woodland, Spike MacPhee, Desdemona Enfield, Mic Bowman, David Levine, Jeff Ames, Adam Johnson, Kat Prawl, Troy McConaghy, Stephanie Smith, Elizabeth Van Horn, Paul Doherty, Will Scotti, Shenlei Winkler, Alan Boyle, Leslie Maxfield, and many others.